\documentclass[aps,showpacs,preprintnumbers,amsmath,amssymb]{revtex4}

\usepackage{float}
\usepackage{graphics,epsfig}
\usepackage{graphicx}
\usepackage{dcolumn}
\usepackage{bm}
\newcommand{\no}{\nonumber}

\begin{document}
\baselineskip=0.8 cm
\title{{\bf Greybody factor for a scalar field coupling to Einstein's tensor}}
\author{Songbai Chen}
\email{csb3752@163.com} \affiliation{Institute of Physics and
Department of Physics, Hunan Normal University,  Changsha, Hunan
410081, P. R. China \\ Key Laboratory of Low Dimensional Quantum
Structures \\ and Quantum Control of Ministry of Education, Hunan
Normal University, Changsha, Hunan 410081, People's Republic of
China}

\author{Jiliang Jing}
\email{jljing@hunnu.edu.cn} \affiliation{Institute of Physics and
Department of Physics, Hunan Normal University,  Changsha, Hunan
410081, P. R. China \\ Key Laboratory of Low Dimensional Quantum
Structures \\ and Quantum Control of Ministry of Education, Hunan
Normal University, Changsha, Hunan 410081, People's Republic of
China}
\begin{abstract}
\baselineskip=0.6 cm
\begin{center}
{\bf Abstract}
\end{center}

We study the greybody factor and Hawking radiation for a scalar
field  coupling to Einstein's tensor in the background of
Reissner-Nordstr\"{o}m black hole in the low-energy approximation.
We find that the presence of the coupling terms modifies the
standard results in the greybody factor and Hawking radiation. Our
results show that both the absorption probability and Hawking
radiation increase with the coupling constant. Moreover, we also
find that for the stronger coupling, the charge of black hole
enhances the absorption probability and Hawking radiation of the
scalar field, which is different from those of ones without coupling
to Einstein's tensor in the black hole spacetime.

\end{abstract}

 \maketitle
\newpage
\section{Introduction}
Scalar field, associated with spin-$0$ particles in quantum field
theory, has been an object of great interest for physicists in the
latest years.  One of the main reasons is that the models with
scalar fields are relatively simple, which allows us to probe the
detailed features of the more complicated physical system. In
cosmology, scalar fields can be considered as candidate to explain
the inflation of the early Universe \cite{inf1} and the accelerated
expansion of the current Universe \cite{1a,2a,3a}. In the Standard
Model of particle physics, the scalar field presents as the Higgs
boson \cite{Hig}, which would help to explain the origin of mass in
the Universe. Moreover, it has been found that scalar field plays
the important roles in other fundamental physical theories, such as,
Jordan-Brans-Dicke theory \cite{bd}, Kaluza-Klein compactification
theory \cite{KK} and superstring theory \cite{dila}, and so on.

In general, the action contained scalar fields in Einstein's theory
of gravity is
\begin{eqnarray}
S=\int d^4x \sqrt{-g}\bigg[\frac{R}{16\pi
G}+\frac{1}{2}\partial_{\mu}\psi\partial^{\mu}\psi+V(\psi)+\xi
R\psi^2\bigg]+S_m,\label{act1}
\end{eqnarray}
where $\psi$, $R$ and $V(\psi)$ are corresponding to scalar field,
Ricci scalar and scalar potential, respectively. The term $\xi
R\psi^2$ represents the coupling between Ricci scalar $R$ and the
scalar field $\psi$. The dynamical behavior of the scalar field in
the theory (\ref{act1}) have been investigated very extensively in
the modern physics including cosmology and black hole physics. The
more general form of the action contained scalar field in other
theories of gravity is
\begin{eqnarray}
S=\int d^4x \sqrt{-g}\bigg[f(\psi, R, R_{\mu\nu}R^{\mu\nu},
R_{\mu\nu\rho\sigma}R^{\mu\nu\rho\sigma}) +K(\psi,
\partial_{\mu}\psi\partial^{\mu}\psi,\nabla^2\psi, R^{\mu\nu}\partial_{\mu}\psi\partial_{\nu}\psi,\cdot\cdot\cdot)+V(\psi)\bigg]+S_m,\label{act2}
\end{eqnarray}
Here $f$ and $K$ are arbitrary functions of the corresponding
variables. Obviously, the more coupling between scalar field and
curvature are considered in these extensive theories. The
non-minimal coupling between the derivative of scalar field and the
spacetime curvature may appear firstly in some Kaluza-Klein theories
\cite{kkc,kkc1,kkc2}. Amendola \cite{AL1} considered the most
general theory of gravity with the Lagrangian linear in the Ricci
scalar, quadratic in $\psi$, in which the coupling terms have the
forms as follows
\begin{eqnarray}
R\partial_{\mu}\psi\partial^{\mu}\psi,\;\;
R_{\mu\nu}\partial^{\mu}\psi\partial^{\nu}\psi,\;\;
R\psi\nabla^2\psi, \;\;
R_{\mu\nu}\psi\partial^{\mu}\psi\partial^{\nu}\psi,\;\;\partial_{\mu}R\partial^{\nu}\psi,
\;\;\nabla^2R\psi.
\end{eqnarray}
And then he studied the dynamical evolution of the scalar field in
the cosmology by considering only the derivative coupling term
$R_{\mu\nu}\partial^{\mu}\psi\partial^{\nu}\psi$ and obtained some
analytical inflationary solutions \cite{AL1} . Capozziello
\textit{et al.} \cite{AL2,AL3} investigated a more general model of
containing coupling terms $R\partial_{\mu}\psi\partial^{\nu}\psi$
and $R_{\mu\nu}\partial^{\mu}\psi\partial^{\nu}\psi$, and found that
the de Sitter spacetime is an attractor solution in the model.
Recently, Daniel and Caldwell \cite{AL4}  obtained the constraints
on the theory with the derivative coupling term of
$R_{\mu\nu}\partial^{\mu}\psi\partial^{\nu}\psi$ by Solar system
tests. In general, a theory with derivative couplings could lead to
that both the Einstein equations and the equation of motion for the
scalar are the fourth-order differential equations. However, Sushkov
\cite{AL5} studied recently the model in which the kinetic term of
the scalar field only coupled with the Einstein tensor and found
that the equation of motion for the scalar field can be reduced to
second-order differential equation. This means that the theory is a
``good" dynamical theory from the point of view of physics. Gao
\cite{g1} investigated the cosmic evolution of a scalar field with
the kinetic term coupling to more than one Einstein tensors and
found the scalar field presents some very interesting characters. He
found that the scalar field behaves exactly as the pressureless
matter if the kinetic term is coupled to one Einstein tensor and
acts nearly as a dynamic cosmological constant if it couples with
more than one Einstein tensors. The similar investigations have been
considered in Refs.\cite{g2,g3}. These results will excite more
efforts to be focused on the study of the scalar field coupled with
tensors in the more general cases.

Since black hole is another fascinating object in modern physics, it
is of interest to extend the study of the properties of the scalar
field when it is kinetically coupled to the Einstein tensors in the
background of a black hole. In this Letter, we will investigate the
greybody factor and Hawking radiation of the scalar field coupling
only to the Einstein tensor $G^{\mu\nu}$ in the
Reissner-Nordstr\"{o}m black hole spacetime by using the matching
technique, which has been widely used in evaluating the absorption
probabilities and Hawking radiations of various black holes
\cite{Kanti,Kan1,Kan2,Kan3,Kan4,Kan5,Kan6,Haw3,Haw4,Haw5}. We find
that the presence of the coupling terms enhances both the absorption
probability and Hawking radiation of the scalar field in the black
hole spacetime. Moreover, we also find that for the stronger
coupling, the absorption probability and Hawking radiation of the
scalar field increase with the charge of the black hole, which is
different from those of scalar one without coupling to Einstein's
tensor.

The Letter is organized as follows: in the following section we will
introduce the action of a scalar field coupling to Einstein's tensor
and derive its master equation in the Reissner-Nordstr\"{o}m black
hole spacetime. In Sec.III, we obtain the expression of the
absorption probability in the low-energy limit by using the matching
technique. In section IV, we will calculate the absorption
probability and the luminosity of Hawking radiation for the coupled
scalar field. Finally, in the last section we will include our
conclusions.

\section{Master equation for a scalar field coupling with Einstein's tensor in the charged black hole spacetime}

Let us consider the action of the scalar field coupling to the
Einstein's tensor $G^{\mu\nu}$ in the curved spacetime \cite{AL5},
\begin{eqnarray}
S=\int d^4x \sqrt{-g}\bigg[\frac{R}{16\pi
G}+\frac{1}{2}\partial_{\mu}\psi\partial^{\mu}\psi+\frac{\eta}{2}G^{\mu\nu}\partial_{\mu}\psi\partial_{\nu}\psi\bigg].\label{acts}
\end{eqnarray}
The coupling between Einstein's tensor $G^{\mu\nu}$ and the scalar
field $\psi$ is represented by the term
$\frac{\eta}{2}G^{\mu\nu}\partial_{\mu}\psi\partial_{\nu}\psi$,
where $\eta$ is coupling constant with dimensions of length-squared.
In general, the presence of such a coupling term brings some effects
to the original metric of the background. However, it is very
difficult for us to obtain an analytic solution for the action
(\ref{acts}). Actually, comparing with the mass of the black hole,
one can find that the energy of a scalar field is very tiny so that
its back-reaction effects on the background can be neglected
exactly. Here, we treat the weak external scalar filed as a probe
field, and then study the effects of the coupling constant $\eta$ on
the greybody factor and Hawking radiation of the scalar filed in the
background of a black hole spacetime.

Varying the action with respect to $\psi$, one can obtain the
modified Klein-Gordon equation
\begin{eqnarray}
\frac{1}{\sqrt{-g}}\partial_{\mu}\bigg[\sqrt{-g}\bigg(g^{\mu\nu}+\eta
G^{\mu\nu}\bigg)\partial_{\nu}\psi\bigg] =0,\label{WE}
\end{eqnarray}
which is a second order differential equation. Obviously, all the
components of the tensor $G^{\mu\nu}$ vanish in the Schwarzschild
black hole spacetime because it is the vacuum solution of the
Einstein's field equation. Thus, we cannot probe the effect of the
coupling term on the greybody factor and Hawking radiation in the
background of a Schwarzschild black hole. The simplest black hole
with the non-zero components of the tensor $G^{\mu\nu}$ is
Reissner-Nordstr\"{o}m one, which can be described by the metric as
follows
\begin{eqnarray}
ds^2&=-&fdt^2+\frac{1}{f}dr^2+r^2
d\theta^2+r^2\sin^2{\theta}d\phi^2,\label{m1}
\end{eqnarray}
with
\begin{eqnarray}
f=1-\frac{2M}{r}+\frac{q^2}{r^2},
\end{eqnarray}
where $M$ is the mass and $q$ is the charge of the black hole. The
Einstein's tensor $G^{\mu\nu}$ for the metric (\ref{m1}) has a form
\begin{eqnarray}
G^{\mu\nu}= \frac{q^2}{r^4} \left(\begin{array}{cccc}
 -\frac{1}{f}&&&\\
 &f&&\\
 &&-\frac{1}{r^2}&\\
 &&&-\frac{1}{r^2\sin^2\theta}
\end{array}\right).
\end{eqnarray}
Adopting to the spherical harmonics
$\psi(t,r,\theta,\phi)=e^{-i\omega t}R(r)Y_{lm}(\theta,\phi)$, we
find that the equation (\ref{WE}) can be separable and the radial
function $R(r)$ obeys to
\begin{eqnarray}
\frac{1}{r^2}\frac{d}{dr}\bigg[r^2\bigg(1+\frac{\eta
q^2}{r^4}\bigg)f\frac{d R(r)}{dr}\bigg] +\bigg[\bigg(1+\frac{\eta
q^2}{r^4}\bigg)\frac{\omega^2}{f}-\bigg(1-\frac{\eta
q^2}{r^4}\bigg)\frac{E_{lm}}{r^2}\bigg]R(r)=0,\label{radial}
\end{eqnarray}
where $E_{lm}=l(l+1)$ is the eigenvalues coming from the angular
equation. Clearly, the radial equation (\ref{radial}) contains the
coupling constant $\eta$, which means that the presence of the
coupling term will change the evolution of the scalar field in the
Reissner-Nordstr\"{o}m black hole spacetime. Moveover, as the charge
$q$ vanishes, it is easy to obtain that the effects of the coupling
constant $\eta$ disappears and the radial equation (\ref{radial})
reduces to that in the Schwarzschild black hole spacetime.

The solution of the radial function $R(r)$ will help us to obtain
the absorption probability $|A_{l}|^2$ and the luminosity of Hawking
radiation for a scalar field coupling to Einstein's tensor in the
Reissner-Nordstr\"{o}m black hole spacetime.

\section{Greybody Factor in the Low-Energy Regime}

In this section, we will present an analytic expressions for the
greybody factors for the emission of a scalar field coupling to
Einstein's tensor in the Reissner-Nordstr\"{o}m black hole
spacetime. Since the radial equation (\ref{radial}) is generally
nonlinear, it is very difficult to obtain its analytic solution.
However, following
Refs.\cite{Kanti,Kan1,Kan2,Kan3,Kan4,Kan5,Kan6,Haw3,Haw4,Haw5}, we
can provide an approximated solution of the radial equation
(\ref{radial}) by employing the matching technique. Firstly, we can
derive the analytic solutions in the near horizon ($r\simeq r_+$)
and far-field $(r\gg r_+)$ regimes in the low-energy limit (i.e.,
$\omega\ll T_H$ and $\omega r_+\ll 1$), respectively. Then, we can
match smoothly these two radial functions $R(r)$ in an intermediate
region since the wave function in physics is continuous everywhere.
Through this matching technique we can construct a smooth analytical
solution of the radial equation valid throughout the entire
spacetime and extract further the ratio between two coefficients
$A^{(\infty)}_{out}$ and $A^{(\infty)}_{in}$ in  Eq. (\ref{rf6}),
which helps us define the greybody factor. The main reason that we
here adopt to the low-energy approximation is that in this limit we
can neglect the back-reaction of a scalar field to the background
metric during the emission process. Moreover, it is well known that
the greybody factor modifies the spectrum of emitted particles from
that of a perfect thermal blackbody. However, in the high-energy
regime, the greybody factor is independent of the energy $\omega$ of
the particle and the spectrum is exactly like that of a perfect
blackbody for every particle species \cite{Haw3}, while in the
low-energy regime, it encodes information about the near horizon
structure of a black hole and about the particles emitted by the
black hole.

Now, we focus on the near-horizon regime and perform the following
transformation of the radial variable as in Refs.
\cite{Haw3,Haw4,Haw5}
\begin{eqnarray}
r\rightarrow f\Rightarrow \frac{d
f}{dr}=(1-f)\;\frac{\mathcal{A}}{r},
\end{eqnarray}
with
\begin{eqnarray}
\mathcal{A}=1-\frac{q^2}{2Mr-q^2}.\label{ax}
\end{eqnarray}
The equation (\ref{radial}) near the horizon $(r\sim r_+)$ can be
rewritten as
\begin{eqnarray}
f(1-f)\frac{d^2R(f)}{d f^2}+(1-D_*f)\frac{d R(f)}{d f}
+\bigg[\frac{K^2_*}{\mathcal{A}(r_+)^2(1-f)f}
-\frac{E_{lm}}{\mathcal{A}(r_+)^2(1-f)}\bigg(\frac{r^4_+-\eta
q^2}{r^4_++\eta q^2}\bigg)\bigg]R(f)=0,\label{r1}
\end{eqnarray}
where
\begin{eqnarray}
K_*=\omega r_+, \;\;\;\;\;\;D_*=1-\frac{2q^2r^2_+(r^4_+-2\eta
r^2_++3\eta q^2)}{(r^2_+-q^2)^2(r^4_++\eta q^2)}.\label{kx}
\end{eqnarray}
Making the field redefinition $R(f)=f^{\alpha}(1-f)^{\beta}F(f)$,
one can find that the equation (\ref{r1}) can be rewritten as a form
of the hypergeometric equation
\begin{eqnarray}
f(1-f)\frac{d^2F(f)}{d f^2}+[c-(1+a+b)f]\frac{d F(f)}{d f}-ab
F(f)=0,\label{r2}
\end{eqnarray}
with
\begin{eqnarray}
a=\alpha+\beta+D_*-1,\;\;\;\;\;\;\;\;\;\;
b=\alpha+\beta,\;\;\;\;\;\;\;\;\;\;\;\;\; c=1+2\alpha.
\end{eqnarray}
Considering the constraint coming from coefficient of $F(f)$, one
can easy to obtain that the power coefficients $\alpha$ and $\beta$
satisfy
\begin{eqnarray}
\alpha^2+\frac{K^2_*}{\mathcal{A}(r_+)^2}=0,
\end{eqnarray}
and
\begin{eqnarray}
\beta^2+\beta(D_*-2)+\frac{1}{\mathcal{A}(r_+)^2}\bigg[K^2_*-
E_{lm}\bigg(\frac{r^4_+-\eta q^2}{r^4_++\eta q^2}\bigg)\bigg]=0,
\end{eqnarray}
respectively. These two equations admit that the parameters $\alpha$
and $\beta$ have the forms
\begin{eqnarray}
&&\alpha_{\pm}=\pm \frac{iK_*}{\mathcal{A}(r_+)},\\
&&\beta_{\pm}=\frac{1}{2}\bigg\{(2-D_*)\pm\sqrt{(D_*-2)^2-\frac{4}{\mathcal{A}(r_+)^2}\bigg[K^2_*-
E_{lm}\bigg(\frac{r^4_+-\eta q^2}{r^4_++\eta q^2}\bigg)\bigg]}
\;\bigg\},\label{bet}
\end{eqnarray}
Following the operation in ref.\cite{Haw3,Haw4,Haw5} and using the
boundary condition that no outgoing mode exists near the horizon, we
can obtain that the parameters $\alpha=\alpha_-$ and
$\beta=\beta_-$. Thus the asymptotic solution near horizon has the
form
\begin{eqnarray}
R_{NH}(f)=A_-f^{\alpha}(1-f)^{\beta}F(a, b, c; f),
\end{eqnarray}
where $A_{-}$ is an arbitrary constant.

Let us now to stretch smoothly the near horizon solution to the
intermediate zone. As done in ref.\cite{Haw3,Haw4,Haw5}, we can make
use of the property of the hypergeometric function \cite{mb} and
change its argument in the near horizon solution from $f$ to $1-f$
\begin{eqnarray}
R_{NH}(f)&=&A_-f^{\alpha}(1-f)^{\beta}\bigg[\frac{\Gamma(c)\Gamma(c-a-b)}{\Gamma(c-a)\Gamma(c-b)}
F(a, b, a+b-c+1; 1-f)\nonumber\\
&+&(1-f)^{c-a-b}\frac{\Gamma(c)\Gamma(a+b-c)}{\Gamma(a)\Gamma(b)}
F(c-a, c-b, c-a-b+1; 1-f)\bigg].\label{r2}
\end{eqnarray}
As $r\gg r_+$, the function $(1-f)$ can be approximated as
\begin{eqnarray}
1-f=\frac{2Mr-q^2}{r^2}\simeq \frac{2M}{r},
\end{eqnarray}
and then the near horizon solution (\ref{r2}) can be simplified
further to
\begin{eqnarray}
R_{NH}(r)\simeq C_1r^{-\beta}+C_2r^{\beta+D_*-2}\label{rn2},
\end{eqnarray}
with
\begin{eqnarray}
C_1=A_-(2M)^{\beta}
\frac{\Gamma(c)\Gamma(c-a-b)}{\Gamma(c-a)\Gamma(c-b)},\label{rn3}
\end{eqnarray}
\begin{eqnarray}
C_2=A_-(2M)^{-(\beta+D_*-2)}\frac{\Gamma(c)\Gamma(a+b-c)}{\Gamma(a)\Gamma(b)}.\label{rn4}
\end{eqnarray}
In order to obtain a solution in the far field region, we expand the
wave equation (\ref{radial}) as a power series in $1/r$ and keep
only the leading terms
\begin{eqnarray}
\frac{d^2R_{FF}(r)}{dr^2}+\frac{2}{r}\frac{dR_{FF}(r)}{d
r}+\bigg(\omega^2-\frac{E_{lm}}{r^2}\bigg)R_{FF}(r)=0.
\end{eqnarray}
This is usual Bessel equation. Thus the solution of radial master
equation (\ref{radial}) in the far-field limit can be expressed as
\begin{eqnarray}
R_{FF}(r)=\frac{1}{\sqrt{r}}\bigg[B_1J_{\nu}(\omega\;r)+B_2Y_{\nu}
(\omega\;r)\bigg],\label{rf}
\end{eqnarray}
where $J_{\nu}(\omega\;r)$ and $Y_{\nu}(\omega\;r)$ are the first
and second kind Bessel functions, $\nu=l+\frac{1}{2}$. $B_1$ and
$B_2$ are integration constants. In order to stretch the far-field
solution (\ref{rf}) towards small radial coordinate, we take the
limit $r\rightarrow 0$ and obtain
\begin{eqnarray}
R_{FF}(r)\simeq\frac{B_1(\frac{\omega\;r}{2})^{\nu}}{\sqrt{r}\;\Gamma(\nu+1)}
-\frac{B_2\Gamma(\nu)}{\pi
\sqrt{r}\;(\frac{\omega\;r}{2})^{\nu}}.\label{rfn2}
\end{eqnarray}
In the low-energy limit, the two power coefficients in
Eq.(\ref{rn2}) can be approximated as
\begin{eqnarray}
-\beta &\simeq &l + O(\omega^2),\no\\
(\beta+D_*-2)&\simeq &-(l+1)+ O(\omega^2),
\end{eqnarray}
respectively. By matching the corresponding coefficients between
Eqs.(\ref{rn2}) and (\ref{rfn2}), we can obtain two relations
between $C_1,\;C_2$ and $B_1,\;B_2$. Removing $A_-$, we can obtain
the ratio between the coefficients $B_1,\; B_2$
\begin{eqnarray}
B\equiv\frac{B_1}{B_2}&=&-\frac{1}{\pi}\bigg[\frac{1}{\omega\;M}\bigg]^{2l+1}
\bigg(\frac{2l+1}{2}\bigg)\Gamma^2(l+\frac{1}{2})
\nonumber\\
&\times&\;\frac{
\Gamma(c-a-b)\Gamma(a)\Gamma(b)}{\Gamma(a+b-c)\Gamma(c-a)\Gamma(c-b)}.
\label{BB}
\end{eqnarray}
In the asymptotic region $r\rightarrow \infty$, the solution in the
far-field can be expressed as
\begin{eqnarray}
R_{FF}(r)&\simeq &
\frac{B_1+iB_2}{\sqrt{2\pi\;\omega}\;r}e^{-i\omega\;r}+
\frac{B_1-iB_2}{\sqrt{2\pi\;\omega}\;r}e^{i\omega\;r}\no\\
&=& A^{(\infty)}_{in}\frac{e^{-i\omega\;r}}{r}
+A^{(\infty)}_{out}\frac{e^{i\omega\;r}}{r}.\label{rf6}
\end{eqnarray}
The absorption probability can be calculated by
\begin{eqnarray}
|\mathcal{A}_{l}|^2=1-\bigg|\frac{A^{(\infty)}_{out}}{A^{(\infty)}_{in}}\bigg|^2
=1-\bigg|\frac{B-i}{B+i}\bigg|^2=\frac{2i(B^*-B)}{BB^*+i
(B^*-B)+1}.\label{GFA}
\end{eqnarray}
Inserting the expression of $B$ (\ref{BB}) into Eq.(\ref{GFA}), we
can probe the properties of absorption probability for the scalar
field coupled with Einstein's tensor in the charged black hole
spacetime in the low-energy limit.

\section{The absorption probability and Hawking radiation of scalar field coupling to Einstein's
tensor}

We are now in a position to calculate the absorption probability and
discuss Hawking radiation of a scalar field coupling to Einstein's
tensor in the background of a Reissner-Nordstr\"{o}m black hole.

In Fig.1, we fixed the coupling constant $\eta$ and plotted the
change of the absorption probability of a scalar particle with the
charge $q$ for the first partial waves ($l=0$) in the
Reissner-Nordstr\"{o}m black hole.  One can easily see that with for
the smaller $\eta$ the absorption probability $A_{l=0}$ decreases
with the charge $q$ of black hole, which is similar to that for the
usual scalar field without coupling to Einstein's tensor. However,
for the larger $\eta$, the absorption probability $A_{l=0}$
increases as the charge $q$ increase, which means that the stronger
coupling between the scalar field and Einstein's tensor changes the
properties of the absorption probability of scalar field in the
black hole spacetime. This phenomenon has not been observed
elsewhere. From Fig.2, we also find that the absorption probability
increases with the increase of the coupling constant $\eta$ for
fixed values of charge $q=0.2$.  These results about the absorption
probability hold true for other values of $l$. It is also shown in
Figs. 3 and 4, in which we plotted the dependence of the absorption
probability on the angular index. Moreover, we see the suppression
of $|A_l|^2$ as the values of the angular index increase. This can
be explained by a fact that the absorption probability
$|A_l|^2\approx \omega^{2l+2}$ in the low-energy approximation
$\omega  r_+\ll1$ \cite{Haw3}, which means that the first partial
wave dominates over all others in the absorption probability. It is
similar to that of the scalar field without coupling to Einstein's
tensor as shown in
Refs.\cite{Kanti,Kan1,Kan2,Kan3,Kan4,Kan5,Kan6,Haw3,Haw4,Haw5}.

\begin{figure}[ht]\label{fig1}
\begin{center}
\includegraphics[width=8.0cm]{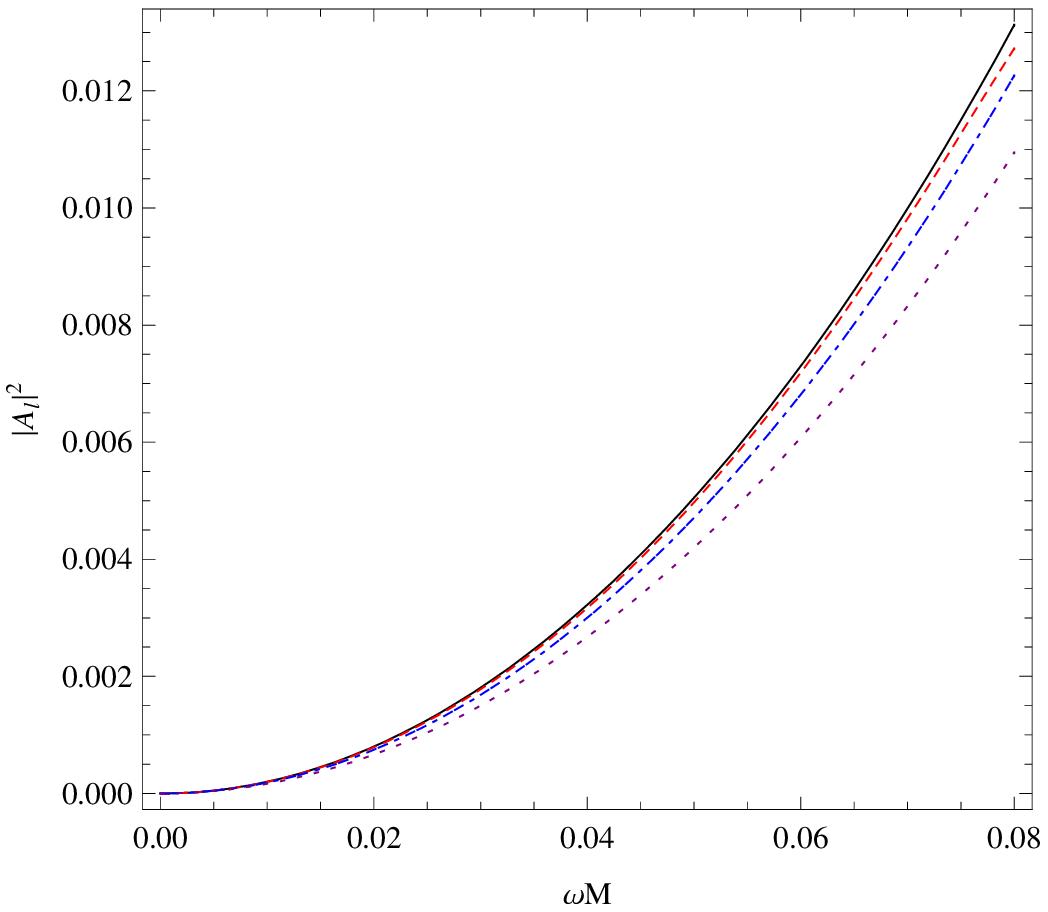}\;\;\;\;\includegraphics[width=8.0cm]{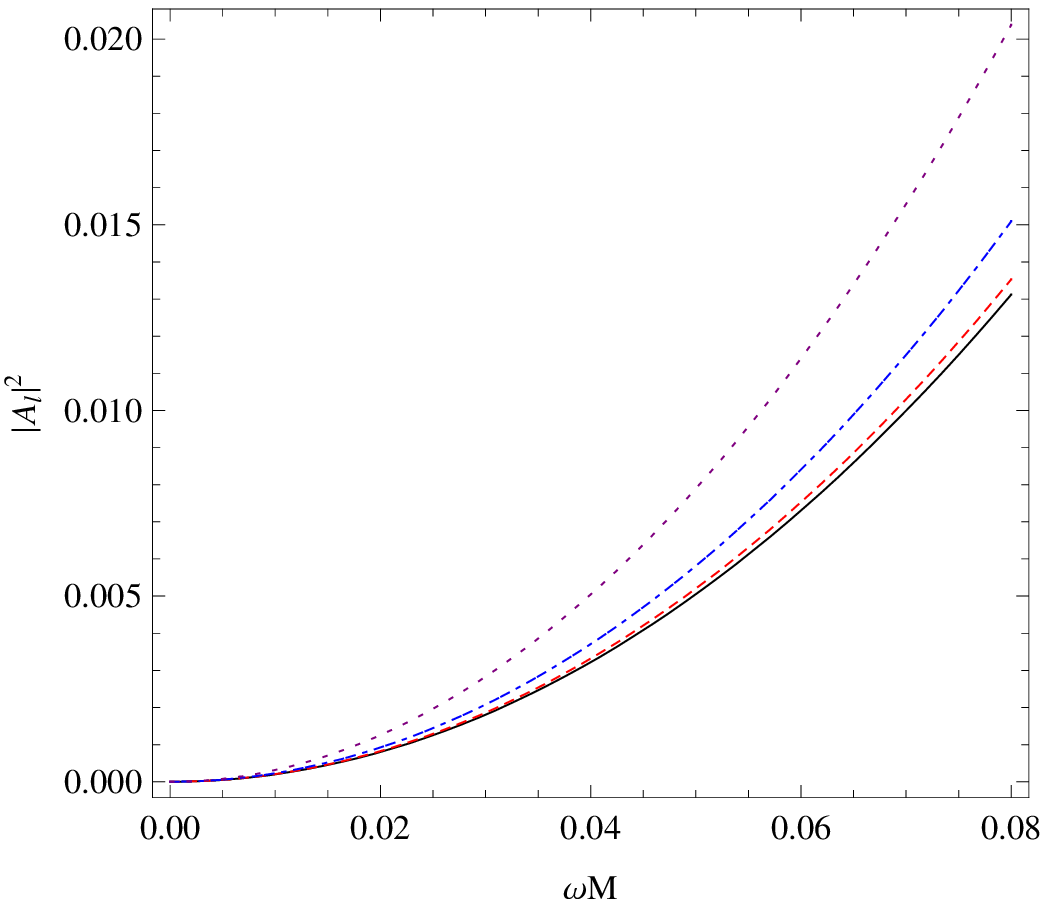}
\caption{Variety of the absorption probability $|A_{l}|^2$ of a
 scalar field with the charge $q$ in the Reissner-Nordstr\"{o}m black hole for
 fixed $l=0$.  The coupling constant $\eta$ is set by $\eta=0.1$ in the left and by  $\eta=1.2$ in the right.
 The solid, dashed, dash-dotted and dotted lines are corresponding to the cases with $q=0,\;0.1,\;0.2,\;0.3$, respectively. We set $2M=1$.}
\end{center}
\end{figure}

\begin{figure}[ht]\label{fig2}
\begin{center}
\includegraphics[width=8.0cm]{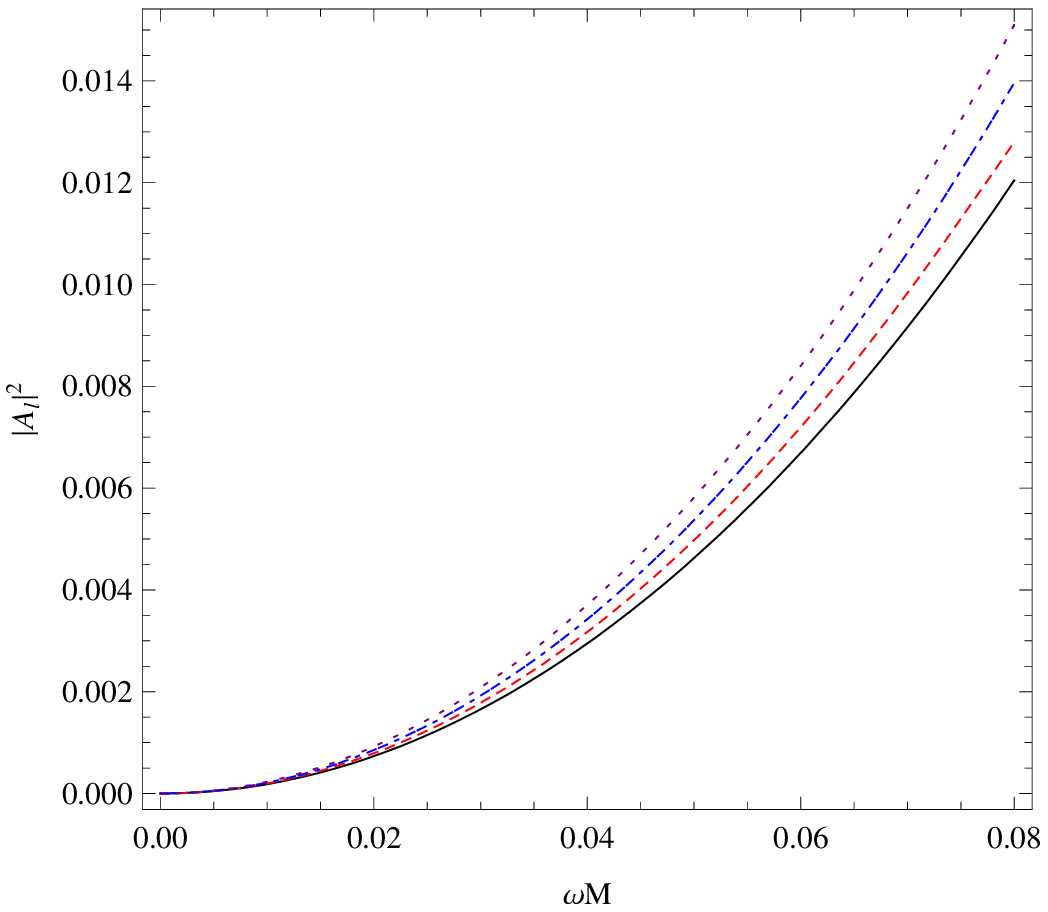}
\caption{The dependence of the absorption probability $|A_{l}|^2$ of
a scalar field on the coupling constant $\eta$ in the
Reissner-Nordstr\"{o}m black hole for fixed $l=0$ and $q=0.2$. The
solid, dashed, dash-dotted and dotted lines are corresponding to the
cases with $\eta=0,\;0.4,\;0.8,\;1.2$, respectively. We set $2M=1$.}
 \end{center}
\end{figure}

\begin{figure}[ht]\label{fig3}
\begin{center}
\includegraphics[width=8.0cm]{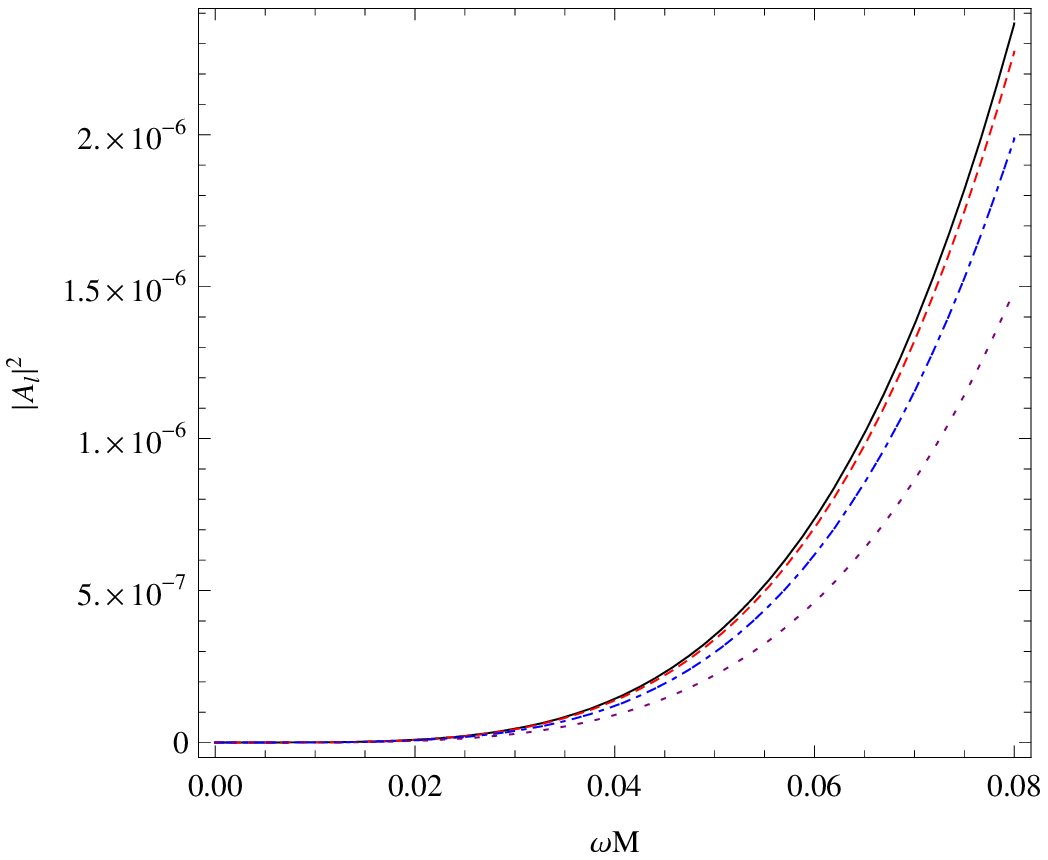}\;\;\;\;\includegraphics[width=8.0cm]{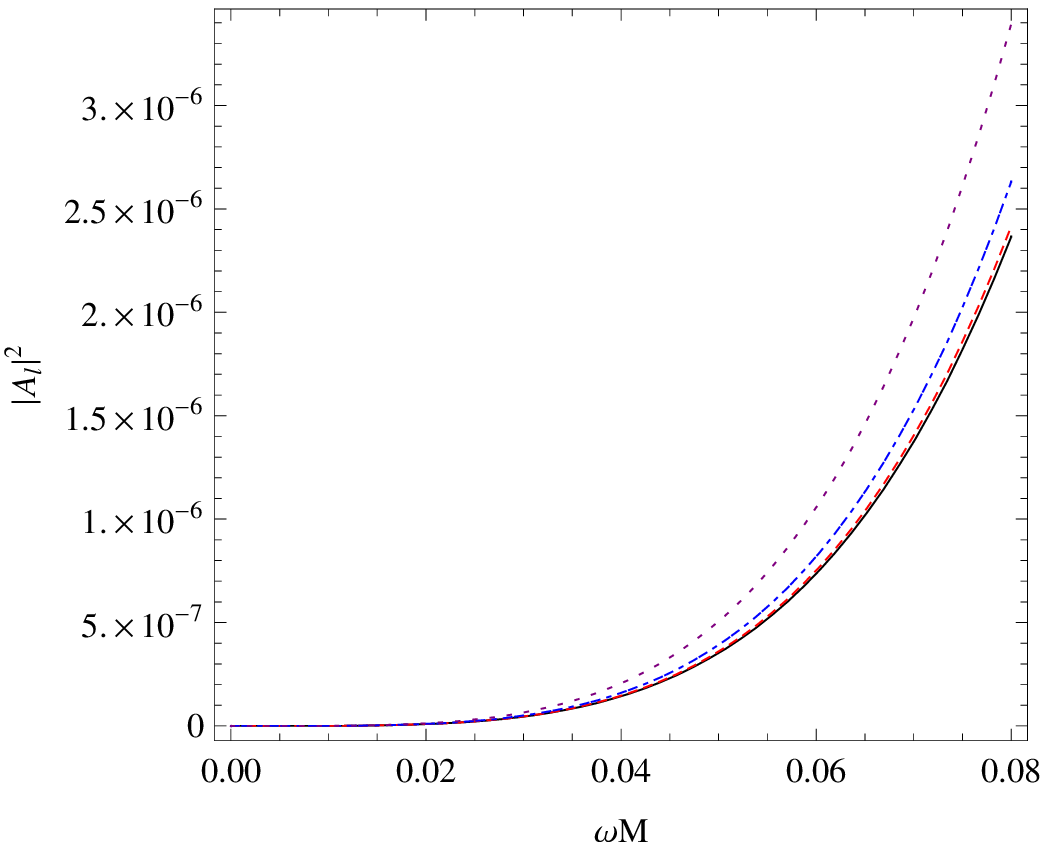}
\caption{Variety of the absorption probability $|A_{l}|^2$ of a
 scalar field with the charge $q$ in the Reissner-Nordstr\"{o}m black hole for
 fixed $l=1$.  The coupling constant $\eta$ is set by $\eta=0.1$ in the left and by  $\eta=1.2$ in the right.
 The solid, dashed, dash-dotted and dotted lines are corresponding to the cases with $q=0,\;0.1,\;0.2,\;0.3$, respectively. We set $2M=1$.}
\end{center}
\end{figure}

\begin{figure}[ht]\label{fig4}
\begin{center}
\includegraphics[width=8.0cm]{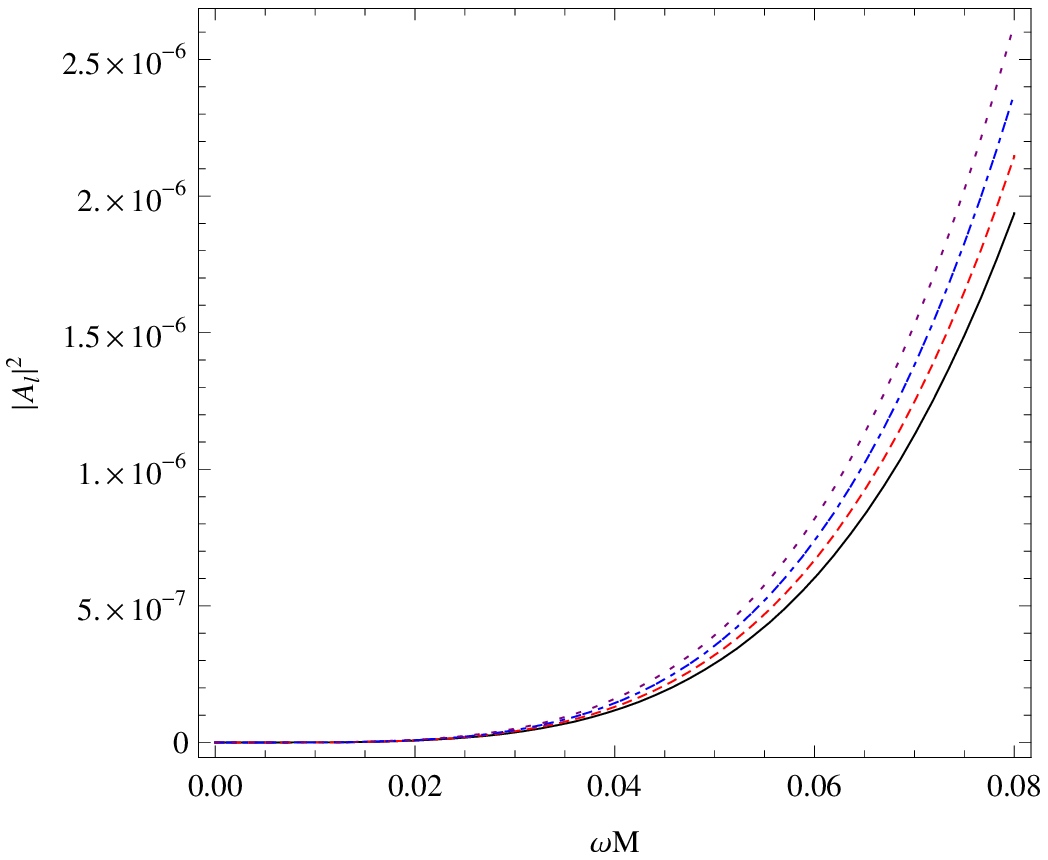}
\caption{The dependence of the absorption probability $|A_{l}|^2$ of
a scalar field on the coupling constant $\eta$ in the
Reissner-Nordstr\"{o}m black hole for fixed $l=1$ and $q=0.2$. The
solid, dashed, dash-dotted and dotted lines are corresponding to the
cases with $\eta=0,\;0.4,\;0.8,\;1.2$, respectively. We set $2M=1$}
\end{center}
\end{figure}

Now let us turn to study the luminosity of the Hawking radiation of
a scalar field coupling to Einstein's tensor in the background of a
Reissner-Nordstr\"{o}m black hole. From above discussion, we know
that in the low-energy approximation $\omega  r_+\ll1$ the mode
$l=0$ plays a dominant role in the greybody factor, which means that
in the scalar emission the major contribution comes from the mode
$l=0$ since the power emission spectra of a scalar particle is in
direct proportion to $|A_{l}|^2$. Thus, we here consider only the
zero mode ($l=0$) and study the effect of the coupling constant
$\eta$ on the luminosity of the Hawking radiation in the background
spacetime. Performing an analysis similar to that in
\cite{Haw3,Haw4,Haw5}, we can obtain that the greybody factor
(\ref{GFA}) in the low-energy limit has a form
\begin{eqnarray}
|A_{l=0}|^2&\simeq& \frac{4\omega^2
r^2_+}{\mathcal{A}(r_+)(2-D_{\ast})}.\label{GFA1}
\end{eqnarray}
Combining it with Hawking temperature $T_H$ of
Reissner-Nordstr\"{o}m  black hole, the luminosity of the Hawking
radiation for the scalar field with coupling to Einstein's tensor is
given by
\begin{eqnarray}
L&=&\int^{\infty}_0\frac{d\omega}{2\pi}
|A_{l=0}|^2\frac{\omega}{e^{\;\omega/T_{H}}-1} \label{LHK}.
\end{eqnarray}
The integral expressions above are just for the sake of completeness
by writing the integral range from $0$ to infinity. However, as our
analysis has focused only in the low-energy regime of the spectrum,
an upper cutoff will be imposed on the energy parameter so that the
low-energy conditions $\omega\ll T_H$ and $\omega r_+\ll 1$ are
satisfied. Like in the radiation of black body, the contribution
from the particles with higher frequencies is very tiny in the power
spectra of Hawking radiation. Thus, the luminosity of the Hawking
radiation for the mode $l=0$ can be approximated as
\begin{eqnarray}
L\approx\frac{2\pi^3}{15}GT^4_H,\label{lh1}
\end{eqnarray}
with
\begin{eqnarray}
G=\frac{r^4_+(r^2_+-q^2)(r^4_++\eta q^2)}{(r^2_+-q^2)^2(r^4_++\eta
q^2)+2q^2r^2_+ (r^4_++3\eta q^2-2\eta r^2_+)}. \label{gs}
\end{eqnarray}
\begin{figure}[ht]\label{fig5}
\begin{center}
\includegraphics[width=8.0cm]{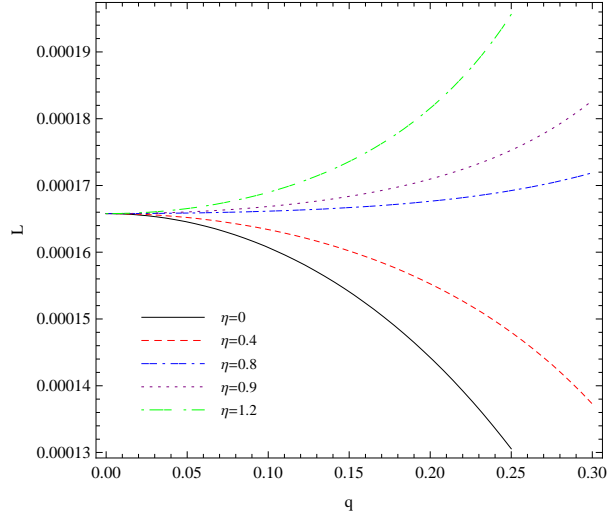}
\caption{Variety of the luminosity of Hawking radiation $L$ of
scalar particles with  the charge $q$ in the Reissner-Nordstr\"{o}m
black hole for fixed $l=0$ and different values of $\eta$.  We set
$2M=1$}
\end{center}
\end{figure}
\begin{figure}[ht]\label{fig6}
\begin{center}
\includegraphics[width=8.0cm]{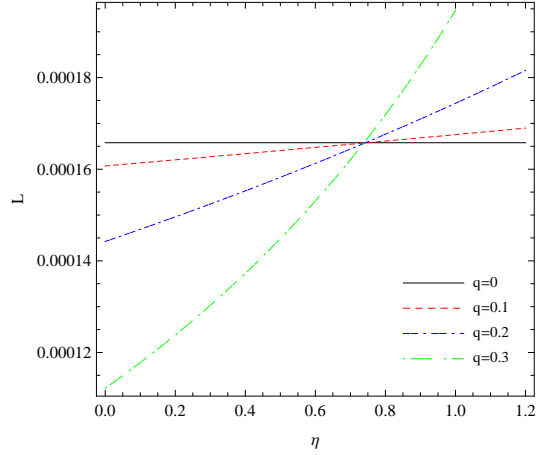}
\caption{The dependence of the luminosity of Hawking radiation $L$
of a scalar field on the coupling constant $\eta$ in the
Reissner-Nordstr\"{o}m black hole for fixed $l=0$ and different
values of $q$. We set $2M=1$.}
\end{center}
\end{figure}
In Fig.5 and 6, we show the dependence of the luminosity of Hawking
radiation (\ref{lh1}) on the charge $q$ and the coupling constant
$\eta$, respectively. From Fig.5, one can easily obtain that with
increase of $q$ the luminosity of Hawking radiation $L$ decreases
for the smaller $\eta$ and increases for the larger $\eta$, which is
similar to the behavior of the absorption probability discussed
previously. The mathematical reason is that the derivative $dL/dq$
has a form
\begin{eqnarray}
\frac{dL}{dq}&=&-\frac{q(r^2_+-q^2)^3}{960\pi
r^2_+[r^4_+(r^4_++q^4)-q^2\eta
(3r^4_+-4q^2r^2_+-q^4)]^2}\bigg[r^{8}_+(3r^6_++9r^4_+q^2+3r^2_+q^2+5q^6)\no \\
&&-2\eta\;r^4_+(2r^8_++3q^2r^6_+-11q^4r^4_+-9q^6r^2_+-5q^8)-q^{4}\eta^2(
5r^6_++7q^2r^4_+-27q^4r^2_++5q^6)\bigg].
\end{eqnarray}
For the smaller $\eta$, we can neglect the term contained the
parameter $\eta$ and find the derivative $dL/dq<0$. For the larger
$\eta$, the derivative $dL/dq$ is dominated by the second term in
the big square-bracket. This leads to that $dL/dq>0$ and we see in
Fig.5 that when $\eta$ is larger the luminosity of Hawking radiation
increases with the charge $q$ of the black hole. In terms of Fig.6,
we have that the luminosity of Hawking radiation $L$ increases
monotonously with the coupling constant $\eta$ for the all $q$.
Similarly, this effect can be explained by a fact the derivative
$dL/d\eta\propto dG/d\eta$ since that the Hawking temperature $T_H$
is independent of the coupling constant $\eta$. From Eq.(\ref{gs}),
we have
\begin{eqnarray}
\frac{dG}{d\eta}=\frac{4r^{10}_+q^2(r^2_+-q^2)^2}{[r^4_+(r^4_++q^4)-q^2\eta
(3r^4_+-4q^2r^2_+-q^4)]^2}>0.
\end{eqnarray}
This means that $G$ increases with the increase of $\eta$, which
leads to the stronger Hawking radiation.

\section{summary}
In this Letter, we have studied the greybody factor and Hawking
radiation for a scalar field coupling to Einstein's tensor in the
background of Reissner-Nordstr\"{o}m black hole in the low-energy
approximation. We have found that the absorption probability and
Hawking radiation depend on the coupling between the scalar field
and Einstein's tensor. The presence of the coupling enhances  both
the absorption probability and Hawking radiation of the scalar field
in the black hole spacetime. Moreover, for the weaker coupling, we
also find that the absorption probability and Hawking radiation
decreases with the charge $q$ of the black hole. It is similar to
that of the scalar field without coupling to Einstein's tensor.
However, for the stronger coupling, the charge $q$ enhances the
absorption probability and Hawking radiation of the black hole,
which could provide a way to detect whether there exist a coupling
between the scalar field and Einstein's tensor or not. It would be
of interest to generalize our study to other black hole spacetimes,
such as rotating black holes etc. Work in this direction will be
reported in the future.

\begin{acknowledgments}

This work was  partially supported by the National Natural Science
Foundation of China under Grant No.10875041,  the Program for
Changjiang Scholars and Innovative Research Team in University
(PCSIRT, No. IRT0964) and the construct program of key disciplines
in Hunan Province. J. Jing's work was partially supported by the
National Natural Science Foundation of China under Grant
No.10675045, No.10875040 and No.10935013; 973 Program Grant No.
2010CB833004 and the Hunan Provincial Natural Science Foundation of
China under Grant No.08JJ3010.
\end{acknowledgments}

\end{document}